\documentclass[prd,aps,tightenlines,preprintnumbers,amsmath,amssymb,showpacs]{revtex4}
\usepackage{graphicx}
\usepackage{dcolumn}
\usepackage{bm}
\usepackage{hyperref}
\usepackage{amsmath}

\def \beq{\begin{equation}}
\def \eeq{\end{equation}}

\def\lsim{\mathrel{\rlap{\lower4pt\hbox{\hskip1pt$\sim$}}
    \raise1pt\hbox{$<$}}}                
\def\gsim{\mathrel{\rlap{\lower4pt\hbox{\hskip1pt$\sim$}}
    \raise1pt\hbox{$>$}}}                
\newcommand{\bra}[1]{\langle #1|}
\newcommand{\ket}[1]{|#1 \rangle}

\begin{document}

\preprint{EFI 09-07}

\title{WIMPless Dark Matter and Meson Decays with Missing Energy}
\author{David McKeen}
\email{mckeen@theory.uchicago.edu}
\affiliation{Enrico Fermi Institute and Department of Physics, University of Chicago, 5640 South Ellis Avenue, Chicago, IL 60637}
\date{\today}

\begin{abstract}
WIMPless dark matter \cite{Feng:2008ya} offers an attractive framework in which dark matter can be very light.  We investigate the implications of such scenarios on invisible decays of bottomonium states for dark matter with a mass less than around $5~{\rm GeV}$.  We relate these decays to measurements of nucleon-dark matter elastic scattering.  We also investigate the effect that a coupling to $s$ quarks has on flavor changing $b\to s$ processes involving missing energy.
\end{abstract}

\maketitle


\begin{section}{Introduction}
Numerous cosmological observations require the existence of nonluminous matter that has not yet been observed in the laboratory -- dark matter.  Efforts to explain electroweak symmetry breaking typically introduce as yet unseen weakly-interacting massive particles (WIMPs).  Scenarios which economically relate the dark matter and electroweak symmetry breaking problems by positing a WIMP as comprising dark matter are quite common.  However, it is worth noting that dark matter may not be connected to the weak scale.  WIMPless models \cite{Feng:2008ya} of dark matter offer a natural scenario in which dark matter can have a mass that is not that of typical WIMP.  These models incorporate the minimal supersymmetric standard model (MSSM) with gauge mediated supersymmetry breaking (GMSB) (for a review of GMSB, see \cite{Giudice:1998bp}) and involve a hidden sector that contains the dark matter.  This hidden sector dark matter is unconnected to the weak scale and may have a mass that is quite different in magnitude.   In particular, dark matter that has GeV scale mass can be accommodated.  Dark matter of this scale has been invoked to explain several tantalizing experimental results, notably those of DAMA/LIBRA \cite{Gondolo:2005hh} and SPI/INTEGRAL \cite{Boehm:2003bt,Hooper:2004qf,Picciotto:2004rp}.  Sub-${\rm GeV}$ scale bosons have also been proposed to explain the ATIC \cite{:2008zzr} and PAMELA \cite{Adriani:2008zq} cosmic ray spectra by providing a Sommerfeld enhancement so that the requisite dark matter annihilation cross sections can agree with relic density constraints and to suppress decays to hadronic final states \cite{Cirelli:2008jk}.  In this note, we examine the implications that a particular WIMPless dark matter model has on the decays of $\bar{b}b$ bound states and on direct detection of nucleon-dark matter elastic scattering.

WIMPless dark matter is introduced in Sec. II along with a particular model. Invisible decays of bottomonium states in this model are considered in Sec. III.  The implications for dark matter-nucleon scattering and its relation to invisible bottomonium decays are discussed in Sec. IV.  In Sec. V, we consider flavor changing $b$-$s$ transitions in this model.  In Sec. VI we conclude.
\end{section}


\begin{section}{The Model}
We consider a model introduced in \cite{Feng:2008ya} and briefly review it here.  We have three sectors: the MSSM, a hidden sector, and a so-called secluded sector which breaks supersymmetry (SUSY).  SUSY breaking in the secluded sector is transmitted to the MSSM and hidden sector with GMSB.  In the secluded sector, a chiral field $S$ attains an expectation value, $\langle S\rangle=M+\theta^2 F$ which breaks SUSY.  This breaking is transmitted to the MSSM by a messenger field, $\Phi$, which transforms nontrivially under the SM gauge group and interacts with $S$ through the superpotential $W=h\bar{\Phi}S\Phi$.  The spinor components of $\Phi$ form Dirac fermions with mass $M_m$ and the scalar components have squared masses $M_m^2\pm F_m$ with $M_m=hM$ and $F_m=h F$.  This generates MSSM superpartner masses of order
\begin{align}
m\sim \frac{g^2}{16\pi^2}\frac{F_m}{M_m}=\frac{g^2}{16\pi^2}\frac{F}{M}
\label{eq:ewscale}
\end{align}
where $g$ is the largest relevant gauge coupling.  In general, a stable thermal relic's density is set by its mass and the coupling strength of its annihilation channels, namely, $\Omega\propto m^2/g^4$.  Combining this and Eq.~\ref{eq:ewscale} fixes the density of a stable thermal relic in terms of the components of $\langle S\rangle$, $\Omega\propto m^2/g^4\sim F^2/(16\pi^2 M)^2$.  The WIMP ``miracle" is that we get a relic density of the right order ($\Omega\sim 0.25$) if we use weak scale masses and couplings for $m$ and $g$.

We now repeat this process with a hidden sector.  We introduce a hidden sector messenger, $\Phi_X$, coupled to $S$ through $W_X=h_X\bar{\Phi}_XS\Phi_X$.  As before, the spinor components have mass $M_{mX}$ and the scalars have squared masses $M_{mX}^2\pm F_{mX}$ with $M_{mX}=h_XM$ and $F_{mX}=h_X F$.  This sets the scale of hidden sector masses,
\begin{align}
m_X\sim \frac{g_X^2}{16\pi^2}\frac{F_{mX}}{M_{mX}}=\frac{g_X^2}{16\pi^2}\frac{F}{M}
\label{eq:xscale}
\end{align}
with $g_X$ the largest relevant gauge coupling in the hidden sector.  Then hidden sector thermal relics have a density given by $\Omega_X\propto m_X^2/g_X^4\sim F^2/(16\pi^2 M)^2$.  Thus, we find that a stable hidden thermal relic will have roughly the right density, $\Omega_X\sim 0.25$, since its value is set by the ratio $F/M$ as in the WIMP case -- a WIMPless ``miracle."

A concrete WIMPless model, with a large hidden sector, was shown in detail in \cite{Feng:2008mu} to be able to reproduce the observed dark matter relic density, illustrating that the general formulation above works in practice.

In the standard GMSB scenario, there is a problem with gravitino production in the decay of the dark matter candidate due to the lightness of stable Standard Model (SM) particles.  This lightness is due to extremely suppressed Yukawa couplings and, as in \cite{Feng:2008ya}, we assume that such a situation does not exist in the hidden sector.

It is phenomenologically interesting to consider a case in which the SM is coupled to the hidden sector dark matter through some connector.  These interactions often occur in intersecting brane models in which a connector has both hidden sector and SM gauge quantum numbers \cite{Cvetic:2001nr}.  We consider a case in which the hidden sector dark matter is a scalar, denoted by $X$.  It is coupled to SM fermions, written as $f$ here, through a chiral fermion, $Y$, via the interaction Lagrangian,
\begin{align}
{\cal L}_{\rm int}&=\lambda_f X\bar{Y}_L f_L+\lambda_f X\bar{Y}_R f_R~~~.
\label{eq:lagrangian}
\end{align}
This model allows for dark matter masses in the range $10~{\rm MeV}\lsim m_X \lsim 10~{\rm TeV}$.  The lower bound is set by the requirement that $X$ be in thermal equilibrium during freeze out so that the expression for the relic density of $X$, $\Omega_X\propto m_X^2$, is valid.  The upper bound is found by enforcing that $g_X$ remain perturbative while giving roughly the right relic density.  We are interested here in very light dark matter, $m_X \lsim 5~{\rm GeV}$.  The connector $Y$ is subject to limits on extra colored particles from the Tevatron: $m_Y>258~{\rm GeV}$ \cite{Kribs:2007nz}.  Perturbativity requires $m_Y\lsim 500~{\rm GeV}$ since $Y$ gets contributions to its mass from electroweak symmetry breaking.  The relic density estimate for $X$, $\Omega_X\propto  F^2/(16\pi^2 M)^2$, is not affected as long as $\lambda_f\lsim g_{\rm weak}$.  This Lagrangian generates several interesting signals which we describe below.  
\end{section}


\begin{section}{Invisible $\chi_{b0}$ Decays}
The Lagrangian of Eq.~\ref{eq:lagrangian} induces the reaction $b\bar{b}\to XX$ when we take $f=b$.  Similar interactions were considered in \cite{McElrath:2005bp} involving $J^{PC}=0^{-+},1^{--}$ quarkonium states.  In this case the relevant quarkonium state has $J^{PC}=0^{++}$.  In the limit of large $m_Y$ the $t$-channel (shown in Fig.~\ref{fig:diagram1}) contribution to the amplitude is
\begin{align}
i{\cal M}_{t}(b\bar{b}\to XX)&\simeq-\frac{i\lambda_b^2}{m_Y}\bar{v}(p^\prime)u(p)~~~.
\label{eq:ME_t}
\end{align}
In this limit, the $u$-channel contribution is the same and so the total amplitude is given by
\begin{align}
i{\cal M}(b\bar{b}\to XX)&\simeq-\frac{2i\lambda_b^2}{m_Y}\bar{v}(p^\prime)u(p)~~~.
\label{eq:ME}
\end{align}
\begin{figure}
\begin{center}
\includegraphics{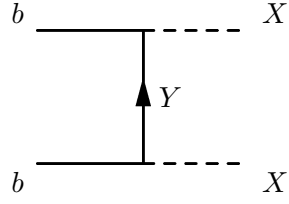}
\caption{$t$-channel $b\bar{b}\to XX$ diagram that leads to $\chi_{b0}\to XX$.}
\label{fig:diagram1}
\end{center}
\end{figure}
In dealing with nonrelativistic bound states it is useful to write the quark spinors in a chiral basis to linear order in their three-momenta,
\begin{align}
u(p)=
\begin{pmatrix}
\sqrt{p\cdot\sigma}\xi\\
\sqrt{p\cdot\bar{\sigma}}\xi
\end{pmatrix}
\simeq\sqrt{m_b}
\begin{pmatrix}
(1-{\bf p}\cdot{\boldsymbol \sigma}/2m_b)\xi\\
(1+{\bf p}\cdot{\boldsymbol \sigma}/2m_b)\xi
\end{pmatrix}~~~,
\label{eq:uspinor}
\end{align}
\begin{align}
v(p^\prime)=
\begin{pmatrix}
\sqrt{p^\prime\cdot\sigma}\xi^\prime\\
-\sqrt{p^\prime\cdot\bar{\sigma}}\xi^\prime
\end{pmatrix}
\simeq\sqrt{m_b}
\begin{pmatrix}
(1+{\bf p}\cdot{\boldsymbol \sigma}/2m_b)\xi^\prime\\
-(1-{\bf p}\cdot{\boldsymbol \sigma}/2m_b)\xi^\prime
\end{pmatrix}~~~,
\label{eq:vspinor}
\end{align}
where $\bf p$ is the center-of-mass momentum of the $b$ quark, $E$ is its energy, and $\xi$ and $\xi^\prime$ are two component spinors of the $b$ and $\bar b$ quarks, respectively. Then,
\begin{align}
i{\cal M}(b\bar{b}\to XX)&\simeq\frac{4i\lambda_b^2}{m_Y}\xi^{\prime\dagger}({\bf p}\cdot{\boldsymbol \sigma})\xi=\frac{4\sqrt{2}i\lambda_b^2}{m_Y}{\bf p}\cdot{\boldsymbol \chi}~~~,
\label{eq:ME2}
\end{align}
where
\begin{align}
{\boldsymbol \chi}={\rm Tr}\left(\frac{{\boldsymbol \sigma}\xi\xi^{\prime\dagger}}{\sqrt{2}}\right)
\end{align}
describes a spin-1 configuration of the quarks.  Thus, this matrix element induces a coupling of $XX$ to the $^3P_0$ $\bar{b}b$ states, the $\chi_{b0}(1P)$ and $\chi_{b0}(2P)$.  These states' wave functions are given by
\begin{align}
\Psi(^3P_0)=\frac{{\boldsymbol \chi}\cdot{\boldsymbol \psi}_{nP}({\bf p})}{\sqrt{3}}~~~,~{\boldsymbol \psi}_{nP}({\bf p})=\int d^3r~{\boldsymbol \psi}_{nP}({\bf r})e^{-i{\bf p}\cdot{\bf r}}~~~, ~{\boldsymbol \psi}_{nP}({\bf r})=\frac{\bf r}{r}\sqrt{\frac{3}{4\pi}}R_{nP}({\bf r})~~~.
\label{eq:wavefn}
\end{align}
Properly accounting for the bound state wave function, the matrix element for the $\chi_{b0}$ to decay to $XX$ is
\begin{align}
{\cal M}(\chi_{b0}(nP)\to XX)&=\sqrt{\frac{2}{M_{\chi_{b0}}}}\int\frac{d^3p}{(2\pi)^3}{\cal M}(b\bar{b}\to XX)\Psi_n({\bf p})
\\&\simeq-\frac{8\lambda_b^2}{\sqrt{3M_{\chi_{b0}}}m_Y}\int\frac{d^3p}{(2\pi)^3}{\bf p}\cdot{\boldsymbol \psi}_{nP}({\bf p})
\\&=\frac{8\lambda_b^2}{\sqrt{3M_{\chi_{b0}}}m_Y}\left[-3i\sqrt{\frac{3}{4\pi}}R_{nP}^\prime(0)\right]\\
&=-\frac{12i}{\sqrt{\pi}}\frac{\lambda_b^2}{\sqrt{M_{\chi_{b0}}}m_Y}R_{nP}^\prime(0)~~~.
\label{eq:ME3}
\end{align}
Thus the decay rate is
\begin{align}
\Gamma(\chi_{b0}(nP)\to XX)&=N_c\left(\frac{1}{2}\right)\left(\frac{1}{2M_{\chi_{b0}}}\right)\left(\frac{\beta}{8\pi}\right)\left|{\cal M}(\chi_{b0}\to XX)\right|^2
\\&\simeq\frac{27}{2\pi^2}\frac{\lambda_b^4\left|R_{nP}^\prime(0)\right|^2}{M_{\chi_{b0}}^2 m_Y^2}\beta
\label{eq:rate}
\end{align}
with $\beta^2=1-4m_X^2/M_{\chi_{b0}}^2$.  Using the expressions for the widths of the $\chi_{b0}$ states \cite{Kwong:1987ak},
\begin{align}
\Gamma(\chi_{b0}(nP)\to \gamma\gamma)&=\frac{\alpha^2\left|R_{nP}^\prime(0)\right|^2}{3 m_b^4}\left(1+0.2 \frac{\alpha_s}{\pi}\right)~~~,
\\
\Gamma(\chi_{b0}(1P)\to gg)&=\frac{6\alpha_s^2\left|R_{1P}^\prime(0)\right|^2}{m_b^4}\left(1+10.0 \frac{\alpha_s}{\pi}\right)~~~,
\\
\Gamma(\chi_{b0}(2P)\to gg)&=\frac{6\alpha_s^2\left|R_{2P}^\prime(0)\right|^2}{m_b^4}\left(1+10.2 \frac{\alpha_s}{\pi}\right)~~~,
\label{eq:rate}
\end{align}
we can find the branching ratios to a pair of $X$ bosons,
\begin{align}
{\cal B}(\chi_{b0}(nP)\to XX)&\simeq 1.3\times 10^{-4}\xi^2~~~,
\label{eq:rate}
\end{align}
with
\begin{align}
\xi=\lambda_b^2\left(\frac{400~{\rm GeV}}{m_Y}\right)\sqrt{\beta}~~~.
\end{align}
$\chi_{b0}$s are produced in radiative decays of the $\Upsilon(2S)$ and $\Upsilon(3S)$, $\Upsilon(2S)\to \gamma \chi_{b0}(1P)$ and $\Upsilon(3S)\to \gamma \chi_{b0}(2P)$.  Thus the $\Upsilon$s will decay to $\gamma XX$ through the $\chi_{b0}$ and we find the branching ratios
\begin{align}
{\cal B}(\Upsilon(2S)\to\gamma XX)&={\cal B}(\Upsilon(2S)\to \gamma \chi_{b0}(1P)){\cal B}(\chi_{b0}(1P)\to XX)
\\
&\simeq\left(4.9\pm0.5\right)\times 10^{-6}\xi^2~~~,
\\
{\cal B}(\Upsilon(3S)\to\gamma XX)&={\cal B}(\Upsilon(3S)\to \gamma \chi_{b0}(2P)){\cal B}(\chi_{b0}(2P)\to XX)
\\
&\simeq\left(7.7\pm0.8\right)\times 10^{-6}\xi^2~~~,
\label{eq:brratio}
\end{align}
where the error shown is only due to that of the measured branching ratios, ${\cal B}(\Upsilon(2S)\to \gamma \chi_{b0}(1P))=\left(3.8\pm0.4\right)\%$ and ${\cal B}(\Upsilon(3S)\to \gamma \chi_{b0}(2P))=\left(5.9\pm0.6\right)\%$ \cite{Amsler:2008zzb}.

A major background for this search at $e^+e^-$ colliders is radiative Bhabha scattering, $e^+ e^-\to\gamma e^+ e^-$, where the electron and positron go undetected.  The cross section is estimated to be (Eq.~\ref{eq:diffxsection})
\begin{align}
\frac{d\sigma}{d\cos\theta}&\simeq\left(\frac{32\alpha^3}{s}\right)\left[\frac{1+\cos^2\theta}{(1-\cos^2\theta)^2}\right] \log\left(\frac{k_{\rm max}}{k_{\rm min}}\right)\log\left[\frac{s^2\left(1-c\right)}{2 m^2 k_{\rm min}k_{\rm max} \left(1+\cos^2\theta\right)}\right]~~~,
\label{eq:diffxsection}
\end{align}
where $\sqrt{s}$ is the center-of-mass energy and $m$ is the mass of the electron.  $c$ is determined by the requirement that the electron and positron are unseen: $\left|\cos\theta_{+,-}\right|>c$ where $\theta_+$ $(\theta_-)$ is the angle the positron (electron) forms with the beam.  The photon has energy between $k_{\rm min}$ and $k_{\rm max}$ and $\theta$ is the photon's angle with the beam direction.

The photon in the $\Upsilon(2S)\to \gamma \chi_{b0}(1P)$ transition will have an energy of $160~{\rm MeV}$ while in the $\Upsilon(3S)\to \gamma \chi_{b0}(2P)$ case it will be $123~{\rm MeV}$.  We have in mind the CLEO experiment and assume a resolution on the photon's energy of 3.5\%, and that $c=0.9$ for the electron and positron to be unseen.  The cross section at both the $\Upsilon(2S)$ and $\Upsilon(3S)$ resonances is then
\begin{align}
\frac{d\sigma}{d\cos\theta}&\simeq\left(75~{\rm pb}\right)\left[\frac{1+\cos^2\theta}{(1-\cos^2\theta)^2}\right] ~~~.
\label{eq:diffxsection2}
\end{align}
The CLEO experiment has collected $1.3~{\rm fb^{-1}}$ on the $\Upsilon(2S)$ resonance and $1.4~{\rm fb^{-1}}$ on the $\Upsilon(3S)$, producing $9.32\times 10^6$ $\Upsilon(2S)$s and $5.88\times 10^6$ $\Upsilon(3S)$s \cite{He:2008xk}.  The signal for $\Upsilon\to \gamma \chi_{b0}\to\gamma XX$ will be distributed as $(3/8)(1+\cos^2\theta)$.  We require $\left|\cos\theta\right|<1/\sqrt{2}$ to cut down on the radiative Bhabha background which is extremely peaked along the beam direction.  We expect this sample to contain
\begin{align}
\left(29\pm 3\right)\xi^2
\end{align}
$\Upsilon(2S)\to \gamma \chi_{b0}(1P)\to\gamma XX$ events and
\begin{align}
\left(28\pm 3\right) \xi^2
\end{align}
$\Upsilon(3S)\to \gamma \chi_{b0}(2P)\to\gamma XX$ events.  Setting these equal to the variation in the background due to radiative Bhabha scattering from Eq.\ \ref{eq:diffxsection2} implies that one can probe
\begin{align}
\xi \gsim 4.3
\end{align}
on the $\Upsilon(2S)$ resonance and
\begin{align}
\xi \gsim 4.4
\end{align}
on the $\Upsilon(3S)$ resonance which correspond to the branching ratios
\begin{align}
{\cal B}(\Upsilon(2S)\to\gamma XX)&\gsim\left(9.1\pm0.5\right)\times 10^{-5} ~~~,
\label{eq:2S_lim}
\\
{\cal B}(\Upsilon(3S)\to\gamma XX)&\gsim\left(15\pm0.5\right)\times 10^{-5}~~~.
\label{eq:3S_lim}
\end{align}
If $m_Y=400~{\rm GeV}$ and $m_X=1~{\rm GeV}$ these limits on $\xi$ imply a limit can be set on the coupling of $Y$ and $X$ to $b$ quarks of $\lambda_b\gsim 2.1$.  This is a little larger than the maximum value of $\lambda_b$ that would not disrupt the relic density estimate of $X$.  We have assumed here an efficiency of 100\%.  The limits on the branching ratios should scale roughly with the inverse of the square root of the efficiency.

Regarding the reach of other $e^+e^-$ collider experiments in setting limits on ${\cal B}(\Upsilon\to\gamma XX)$, we note that the BaBar experiment collected over $100\times 10^6$ $\Upsilon(3S)$ decays \cite{:2008st}.  The reach in the branching ratio, given similar resolutions and efficiencies, scales as $N^{-1/2}$, with $N$ the number of decays, so one would expect BaBar to be able to probe ${\cal B}(\Upsilon(3S)\to\gamma XX)\sim 10^{-5}$.  A Super-B factory like that envisioned by the BELLE collaboration with $10-20~{\rm ab}^{-1}$ collected on the $\Upsilon(4S)$ could collect $10^9$ $\Upsilon(3S)$s and limit  ${\cal B}(\Upsilon(3S)\to\gamma XX)\sim 5\times  10^{-6}$.
\end{section}


\begin{section}{Nucleon-$X$ Scattering}
The Lagrangian of Eq.~\ref{eq:lagrangian} can also lead to nucleon-$X$ interactions like the one shown in Fig.~\ref{fig:diagram2}.  This leads to a spin independent cross section \cite{Feng:2008dz}
\begin{align}
\sigma_{\rm SI}=\frac{\lambda_b^4}{4\pi}\frac{m_N^2\left[ZB_b^p+(A-Z)B_b^n\right]^2}{A^2\left(m_N+m_X\right)^2\left(m_Y-m_X\right)^2}~~~,
\label{eq:xsection}
\end{align}
where $A$ is the atomic mass of the nucleon, $Z$ is its atomic number, and $B_b^{p,n}=\bra{p,n}\bar{b}b\ket{p,n}$.  This matrix element has been calculated using the conformal anomaly \cite{Shifman:1978zn}, and we write  $B_b^{p,n}=(2/27)m_N f_g^{p,n}/m_b$ and, as in \cite{Feng:2008dz}, we take $f_g^p=f_g^n\simeq0.8=f_g$.  Then the cross section becomes
\begin{align}
\sigma_{\rm SI}=\frac{\lambda_b^4}{4\pi}\left(\frac{2}{27}\right)^2\frac{f_g^2 m_N^4}{m_b^2(m_N+m_X)^2(m_Y-m_X)^2}~~~.
\label{eq:xsection}
\end{align} 

\begin{figure}
\begin{center}
\includegraphics{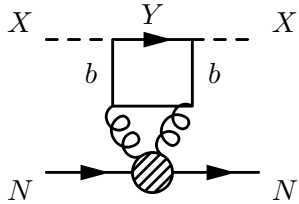}
\caption{Diagram that leads to spin independent $X$-proton cross section.}
\label{fig:diagram2}
\end{center}
\end{figure}

Setting a limit on ${\cal B}(\Upsilon(2S)\to \gamma \chi_{b0}\to\gamma XX)$, through $\lambda_b$, corresponds to setting a limit on $\sigma_{\rm SI}$.  We have plotted in Fig.~\ref{fig:xsection} the reach in $\sigma_{\rm SI}$ that can be obtained for particular limits on ${\cal B}(\Upsilon(2S)\to \gamma \chi_{b0}\to\gamma XX)$.  We use a messenger of mass $m_Y=400~{\rm GeV}$.  The CRESST experiment \cite{Angloher:2002in} sets the most stringent limits on $\sigma_{\rm SI}$ for $m_X\lsim5~{\rm GeV}$.  We have also indicated the range of $\sigma_{\rm SI}$ and $m_X$ that are compatible with the dark matter interpretation of the DAMA/LIBRA data if one allows for dark matter streams \cite{Gondolo:2005hh}.
\begin{figure}
\begin{center}
\rotatebox{270}{\resizebox{80mm}{!}{\includegraphics{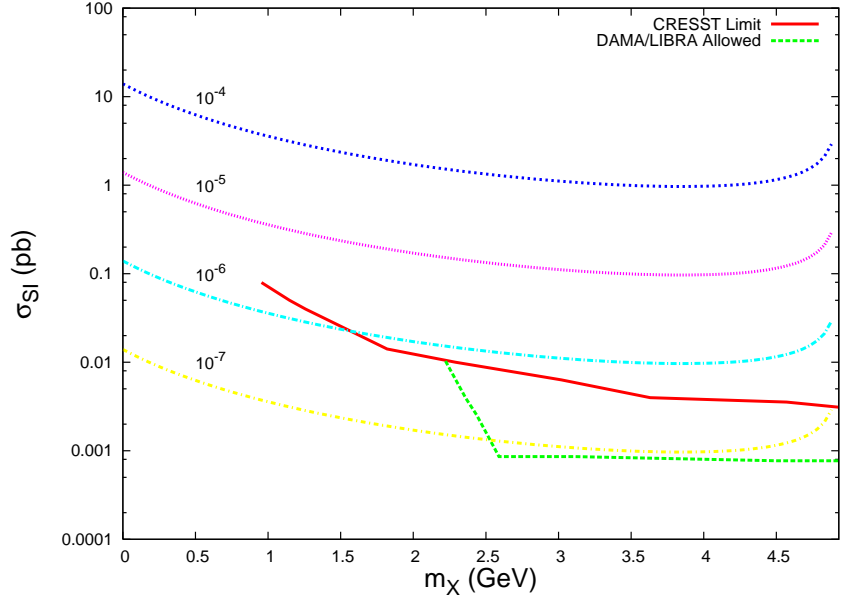}}}
\caption{Spin independent $X$-proton cross section as a function of $m_X$ with contours of constant ${\cal B}(\Upsilon(2S)\to \gamma \chi_{b0}\to\gamma XX)$ indicated.  The CRESST limit \cite{Angloher:2002in} is shown in red (solid) while the area between the red (solid) and green (dashed) lines is consistent with the dark matter interpretation of DAMA/LIBRA if dark matter streams are included \cite{Gondolo:2005hh}.}
\label{fig:xsection}
\end{center}
\end{figure}
However, the reach in $\sigma_{\rm SI}$ for a given limit on ${\cal B}(\Upsilon(2S)\to \gamma \chi_{b0}\to\gamma XX)$ is nearly independent of the value of $m_Y$.  This is because both $\sigma_{\rm SI}$ and ${\cal B}(\Upsilon(2S)\to \gamma \chi_{b0}\to\gamma XX)$, for large $m_Y$, depend only on the quantity $\lambda_b^4/m_Y^2$.

Fig.~\ref{fig:xsection} and the limits from Eqs.~\ref{eq:2S_lim} and ~\ref{eq:3S_lim} show that current CLEO data, in setting a limit on the branching ratio ${\cal B}(\Upsilon\to\gamma XX)\gsim 10^{-4}$, can complement data from direct detection experiments for dark matter particles with a mass $m_X< 1~{\rm GeV}$.  A limit ${\cal B}(\Upsilon\to\gamma XX)\gsim 10^{-6}$ could be possible at a Super-B factory and would be quite competitive with those from direct detection.
\end{section}


\begin{section}{$b$-$s$ Transitions}

The Lagrangian of Eq.~\ref{eq:lagrangian} can give rise to an effective $b$-$s$ quark interaction if both $\lambda_b$ and $\lambda_s$ are nonzero.  If so, to lowest order in $m_Y$, one finds an effective interaction term,
\begin{align}
{\cal L}_{\rm eff}=\frac{2\lambda_b \lambda_s^*}{m_Y}\bar{s}b X^2~~~.
\label{eq:eff_lagrangian}
\end{align}
This will induce the decay $B^+\to K^+ XX$ through the diagram shown in Fig.~\ref{fig:diagram3}.  Using this process to constrain other models of dark matter has been considered in \cite{Bird:2004ts}.  The amplitude for this decay, mediated by the effective Lagrangian of Eq.~\ref{eq:eff_lagrangian}, is
\begin{align}
{\cal M}\left(B^+\to K^+ XX\right)=\frac{8\lambda_b \lambda_s^*}{m_Y}\frac{M_B^2-M_K^2}{m_b-m_s}f_0(q^2)~~~,
\label{eq:eff_lagrangian}
\end{align}
where $q^2=(p_B-p_K)^2$.  $f_0(q^2)$ is estimated, using light-cone sum rules, to be \cite{Ali:1999mm}
\begin{align}
f_0(q^2)=0.3~{\rm Exp}\left[0.63\left(\frac{q^2}{M_B^2}\right)-0.095\left(\frac{q^2}{M_B^2}\right)^2+0.591\left(\frac{q^2}{M_B^2}\right)^3\right]~~.
\end{align}
\begin{figure}
\begin{center}
\includegraphics{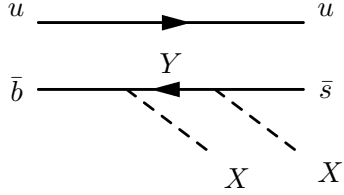}
\caption{Diagram relevant for $B^+\to K^+ XX$.}
\label{fig:diagram3}
\end{center}
\end{figure}
The differential decay rate is then
\begin{align}
\frac{d\Gamma}{dq^2}\left(B^+\to K^+ XX\right)=\frac{1}{8\pi^3 M_B^3}\left|\frac{\lambda_b \lambda_s^*}{m_Y}\right|^2\left(\frac{M_B^2-M_K^2}{m_b-m_s}\right)^2f_0^2(q^2)\lambda^{1/2}(q^2,M_B^2,M_K^2)\sqrt{1-\frac{4m_X^2}{q^2}}
\end{align}
with
\begin{align}\lambda^{1/2}(q^2,M_B^2,M_K^2)=\left[\left(q^2\right)^2+M_B^4+M_K^4-2q^2M_B^2-2q^2M_K^2-2M_B^2M_K^2\right]^{1/2}~~~.
\end{align}
Using $m_b=4.2~{\rm GeV}$ and $m_s=100~{\rm MeV}$, we integrate over $q^2$ and find the branching ratio,
\begin{align}
{\cal B}\left(B^+\to K^+ XX\right)=\left(1.0\times 10^{5}\right)\left|\rho\right|^2 F\left(m_X\right)~~~,
\end{align}
where we define
\begin{align}
\rho=\lambda_b \lambda_s^* \left(\frac{400~{\rm GeV}}{m_Y}\right)
\label{eq:rho}
\end{align}
and $F(m_X)$ describes the allowed phase space as a function of $m_X$,
\begin{align}
F\left(m_X\right)=\frac{\int_{4 m_X^2}^{\left(M_B-M_K\right)^2}dq^2f_0^2(q^2)\lambda^{1/2}(q^2,M_B^2,M_K^2)\sqrt{1-4m_X^2/q^2}}{\int_{0}^{\left(M_B-M_K\right)^2}dq^2f_0^2(q^2)\lambda^{1/2}(q^2,M_B^2,M_K^2)}~~~.
\end{align}
Using the limit ${\cal B}\left(B^+\to K^+ \bar{\nu}\nu\right)<1.4\times 10^{-5}$ \cite{:2007zk}, we can set an upper limit on $\left|\rho\right|$.  For $m_X<2~{\rm GeV}$, we find $\rho<10^{-5}$.  The upper limit on $\rho$ as a function of $m_X$ is shown in Fig.~\ref{fig:rholim}.

\begin{figure}
\begin{center}
\rotatebox{270}{\resizebox{80mm}{!}{\includegraphics{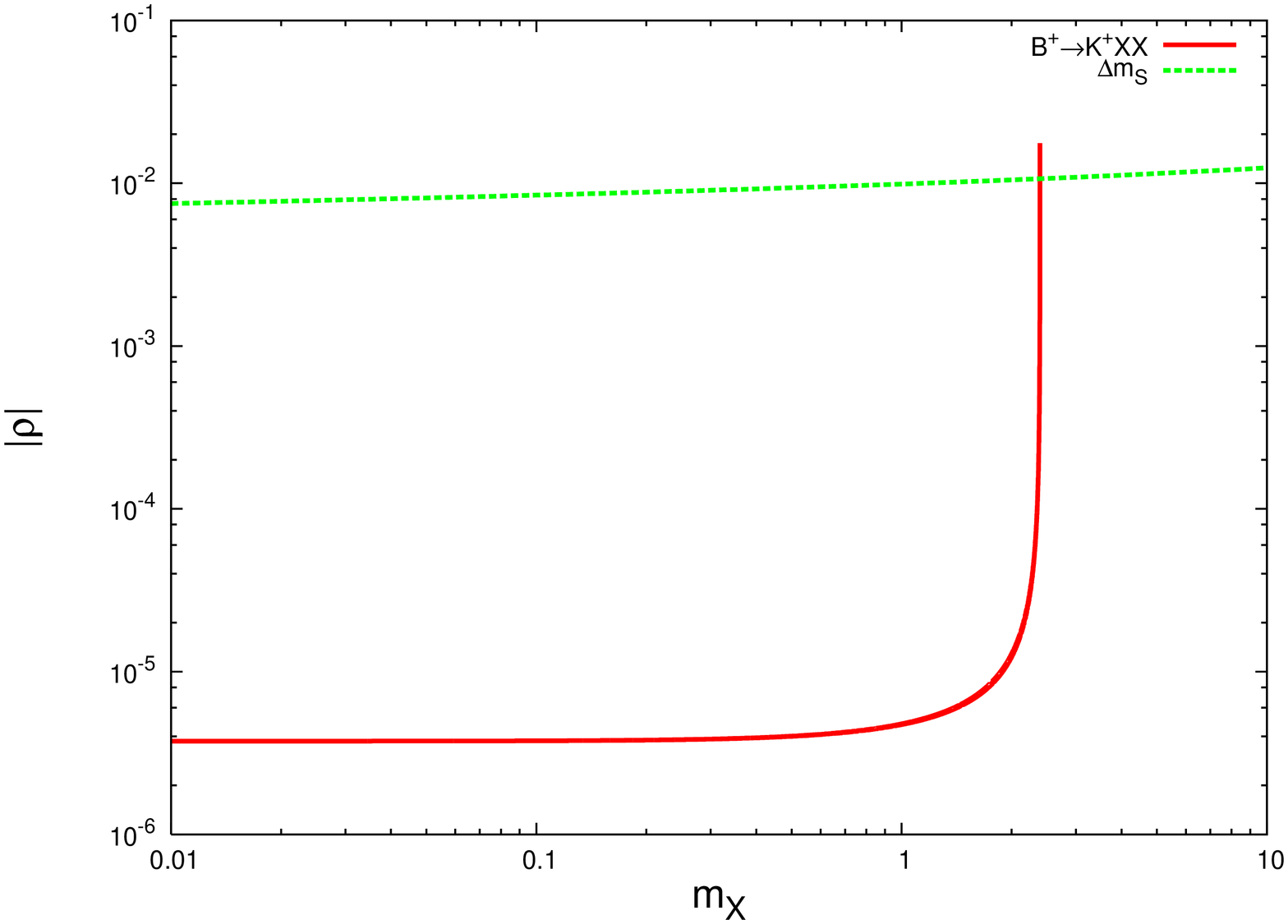}}}
\caption{The upper limit on $\left|\rho\right|$, defined in Eq.\ \ref{eq:rho}, as a function of $m_X$ from ${\cal B}\left(B^+\to K^+ \bar{\nu}\nu\right)<1.4\times 10^{-5}$ \cite{:2007zk} and from the contribution to $\Delta m_s$ from $X$ and $Y$ exchange being less that the measured value of $17.77\pm0.12~{\rm ps}^{-1}$ \cite{Abulencia:2006ze} for $m_Y=400~{\rm GeV}$.}
\label{fig:rholim}
\end{center}
\end{figure}

There is also a contribution to $\bar{B}_s-B_s$ mixing from the Lagrangian of Eq.~\ref{eq:lagrangian} at one-loop through $\bar{b}s\to b\bar{s}$.  One diagram that contributes is shown in Fig.~\ref{fig:diagram4}.
\begin{figure}
\begin{center}
\includegraphics{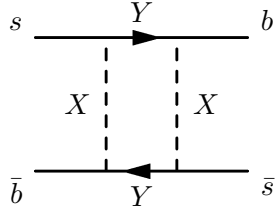}
\caption{One diagram that gives a contribution to $\Delta m_s$.}
\label{fig:diagram4}
\end{center}
\end{figure}
The relevant portion of the amplitude for this is
\begin{align}
{\cal M}\left(\bar{b}s\to b\bar{s}\right)=\frac{\left(\lambda_b \lambda_s^*\right)^2}{32\pi^2 m_Y^2}\left\{\left[g\left(y\right)+h\left(y\right)\right]\bar{b}^\alpha\gamma^5s^\beta\bar{b}^\beta\gamma^5s^\alpha+\left[\frac{g\left(y\right)}{2}-h\left(y\right)\right]\bar{b}^\alpha\gamma^\mu\gamma^5s^\beta\bar{b}^\beta\gamma_\mu\gamma^5s^\alpha\right\}~~~.
\end{align}
where $\alpha$, $\beta$ are color indices, $y=m_X^2/m_Y^2$ and
\begin{align}
g(y)&=\frac{(1-y^2)+2y\log(y)}{(1-y)^3}~~~,
\\
h(y)&=\frac{2(y-1)-(1+y)\log(y)}{(1-y)^3}~~~.
\end{align}
In the limit that $m_X\ll m_Y$, using the vacuum saturation approximation, this yields
\begin{align}
\bra{\bar{B}_s}{\cal M}\ket{B_s}\simeq-\frac{\left(\lambda_b \lambda_s^*\right)^2}{576\pi^2 m_Y^2}\log\left(\frac{m_X^2}{m_Y^2}\right)f_{B_s}^2 M_{B_s}^2\left[8+5\left(\frac{M_{B_s}}{m_b+m_s}\right)^2\right] ~~~,
\end{align}
which gives a contribution to the mass difference
\begin{align}
\Delta m_s&\simeq-\frac{\left(\lambda_b \lambda_s^*\right)^2}{288\pi^2 m_Y^2}\log\left(\frac{m_X^2}{m_Y^2}\right)f_{B_s}^2 M_{B_s}\left[8+5\left(\frac{M_{B_s}}{m_b+m_s}\right)^2\right] 
\\
&\simeq \left(1.82\pm0.12\times 10^{5}~{\rm ps}^{-1}\right)\rho^2\left\{1-0.08\log\left[\left(\frac{m_X}{1~{\rm GeV}}\right)^2\left(\frac{400~{\rm GeV}}{m_Y}\right)^2\right]\right\}~~~,
\end{align}
where we have used $f_{B_s}\simeq 231\pm 15~{\rm MeV}$ \cite{Gamiz:2009ku}.  Conservatively using the total measured value of $\Delta m_s=17.77\pm0.12~{\rm ps}^{-1}$ \cite{Abulencia:2006ze} as an upper limit on the expression above, we can set an upper limit on $\left|\rho\right|$, ignoring any phase difference between $\lambda_b$ and $\lambda_s$.  This limit, for $m_Y=400~{\rm GeV}$, is shown in Fig.~\ref{fig:rholim}.  For values of $m_X$ where $B^+\to K^+XX$ is kinematically allowed, its constraint on $\rho$ is much stronger than that of $\Delta m_s$.

Limits from $b$-$s$ transitions give stricter constraints on couplings of the form of those in Eq.~\ref{eq:lagrangian} than those involving only couplings to $b$ quarks.  We are led to speculate that there may be a Cabibbo-like suppression of the coupling of a heavy chiral fermion, $Y$, to scalar dark matter, $X$, and an $s$ quark relative to that involving a $b$ quark.
\end{section}


\begin{section}{Conclusions}
Bosons with a mass on the $\rm GeV$ scale that are weakly coupled to the SM have been implicated in a number of experimental observations, and may even comprise the dark matter.  WIMPless models of dark matter can naturally accommodate scalar dark matter with a mass in the $\rm GeV$ range.  Scenarios with light dark matter can have many implications for decays of mesons that involve undetected final states (missing energy).  We have seen here an example of a simple WIMPless model involving a fermionic messenger and scalar dark matter that gives novel signatures.  The conclusions drawn here, however, apply to any model involving an interaction of the form of that in Eq.~\ref{eq:lagrangian}.  Existing constraints from collider experiments can verify those from direct detection experiments or, in the case of very light dark matter, be complementary to them.  Comparison of effects involving flavor changing $b$-$s$ transitions and those only involving decay of $\bar{b}b$ states indicates that it is likely that dark matter in this scenario is preferentially coupled to $b$ quarks.  In this case, studies of the decays of bottomonium states involving invisible final state particles provide the opportunity to test these models.
\end{section}


\begin{acknowledgments}
The author would like to thank J. L. Rosner for numerous discussions and helpful suggestions.  This work was supported in part by the United States Department of Energy under Grant No. DE-FG02-90ER40560.
\end{acknowledgments}


\appendix*
\section{Calculation of the Radiative Bhabha Cross Section}

The squared amplitude for radiative Bhabha scattering, in the limit that the electron mass $m=0$, is (Eq. (2.60) of \cite{Berends:1981uq})
\begin{align}
\left|{\cal M}\right|^2&=-\frac{e^6\left(v_q-v_p\right)^2}{s s^\prime t t^\prime}\left[s s^\prime\left(s^2 + s^{\prime2}\right)+t t^\prime\left(t^2 + t^{\prime2}\right)+u u^\prime\left(u^2 + u^{\prime2}\right)\right]~~~,
\end{align}
where
\begin{align}
&s=\left(p_++p_-\right)^2~~~, ~~~s^\prime=\left(q_++q_-\right)^2~~~,
\\
&t=\left(p_+-q_+\right)^2~~~,~~~t^\prime=\left(p_--q_-\right)^2~~~,
\\
&u=\left(p_+-q_-\right)^2~~~,~~~u^\prime=\left(p_--q_+\right)^2~~~,
\\
&v_q=\frac{q_+}{q_+\cdot k}-\frac{q_-}{q_-\cdot k}~~~,~~~v_p=\frac{p_+}{p_+\cdot k}-\frac{p_-}{p_-\cdot k}~~~.
\end{align}
We take $p_-$ ($p_+$) as the four-momenta of the incoming electron (positron), $q_-$ ($q_+$) as the four-momenta of the outgoing electron (positron), and $k$ as the photon's four-momentum.  We write the four-momenta in the center-of-mass frame, neglecting the electron's mass, as
\begin{align}
&p_+=\left(\frac{\sqrt{s}}{2}\right)\left(1,\hat{z}\right)~~~,~~~p_-=\left(\frac{\sqrt{s}}{2}\right)\left(1,-\hat{z}\right)~~~,
\\
&q_+=q_+^0\left(1,\hat{n}_+\right)~~~,~~~q_-=q_-^0\left(1,\hat{n}_-\right)~~~,
\\
&\hat{n}_\pm=\sin\theta_\pm\cos\phi_\pm~\hat{x}+\sin\theta_\pm\sin\phi_\pm~\hat{y}+\cos\theta_\pm~\hat{z}~~~,
\\
&k=\left(\frac{x\sqrt{s}}{2}\right)\left(1,\hat{n}\right)~~~,~~~\hat{n}=\sin\theta\cos\phi~\hat{x}+\sin\theta\sin\phi~\hat{y}+\cos\theta~\hat{z}~~~.
\end{align}
We find
\begin{align}
q_+^0&=\left(\frac{\sqrt{s}}{2}\right)\left[\frac{1-x}{1-(x/2)(1-\hat{n}\cdot\hat{n}_+)}\right]~~~,
\\
q_-^0&=\left(\frac{\sqrt{s}}{2}\right)\left[\frac{1-(x/2)(2-x)(1-\hat{n}\cdot\hat{n}_+)}{1-(x/2)(1-\hat{n}\cdot\hat{n}_+)}\right]~~~,
\\
\hat{n}_-&=-\left[\frac{(1-x)\hat{n}_++x(1-(x/2)(1-\hat{n}\cdot\hat{n}_+))\hat{n}}{1-(x/2)(2-x)(1-\hat{n}\cdot\hat{n}_+)}\right]~~~.
\end{align}
We are interested in cases where the photon's energy is small compared to the center-of-mass energy and the electron and positron are slightly deflected.   That is, we are interested in cases where $x$ is small.  To first order in $x$,
\begin{align}
q_+^0&=\left(\frac{\sqrt{s}}{2}\right)\left[1-\frac{x}{2}(1+\hat{n}\cdot\hat{n}_+)\right]~~~,
\\
q_-^0&=\left(\frac{\sqrt{s}}{2}\right)\left[1-\frac{x}{2}(1-\hat{n}\cdot\hat{n}_+)\right]~~~,
\\
\hat{n}_-&=-(1-x\hat{n}\cdot\hat{n}_+)\hat{n}_+-x\hat{n}~~~.
\end{align}
We use these to find $v_p$ and $v_q$ to lowest order in $x$,
\begin{align}
v_p^0&=\left(\frac{4}{x\sqrt{s}}\right)\frac{\cos\theta}{\sin^2\theta}~~~,
\\
{\bf v}_p&=\left(\frac{4}{x\sqrt{s}}\right)\frac{\hat{z}}{\sin^2\theta}~~~,
\\
v_q^0&=\left(\frac{4}{x\sqrt{s}}\right)\left[\frac{\hat{n}\cdot\hat{n}_+}{1-(\hat{n}\cdot\hat{n}_+)^2}\right]~~~,
\\
{\bf v}_q&=\left(\frac{4}{x\sqrt{s}}\right)\left[\frac{\hat{n_+}}{1-(\hat{n}\cdot\hat{n}_+)^2}\right]~~~.
\end{align}
The dominant contribution to the cross section comes when $\cos\theta_+\simeq1$.  We write $\cos\theta_+=1-a^2/2$ and expand $v_q$ to lowest order in $a$,
\begin{align}
v_q^0&\simeq\left(\frac{4}{x\sqrt{s}}\right)\left[\frac{\cos\theta+a\sin\theta\cos\phi_+(1+2\cot^2\theta)}{\sin^2\theta}\right]~~~,
\\
{\bf v}_q&=\left(\frac{4}{x\sqrt{s}}\right)\left[\frac{\hat{z}+a(\cos\phi_+\hat{x}+\sin\phi_+\hat{y}+2\cot\theta\cos\phi_+\hat{z})}{\sin^2\theta}\right]~~~,
\end{align}
where, without loss of generality, we have set $\cos\phi=1$.  Then we find
\begin{align}
(v_q-v_p)^2&\simeq\left(\frac{16a^2}{x^2s}\right)\left[\frac{\sin^2\theta\cos^2\phi_+(1+2\cot^2\theta)^2-1-4\cot^2\theta\cos^2\phi_+}{\sin^4\theta}\right]
\\
&=-\left(\frac{16a^2}{x^2s}\right)\left[\frac{1-\sin^2\theta\cos^2\phi_+}{\sin^4\theta}\right]~~~.
\end{align}
Now, looking at the other Lorentz invariants that make up the amplitude we also find to lowest order in $x$ and $a$,
\begin{align}
s^\prime&\simeq s
\\
t& \simeq t^\prime\simeq-\left(\frac{s}{2}\right)(1-\cos\theta_+)\simeq -\frac{sa^2}{4}
\label{eq:t}
\\
u& \simeq u^\prime\simeq-s
\end{align}
Collecting all of this we find the squared amplitude to lowest order in $x$ and $a$,
\begin{align}
\left|{\cal M}\right|^2&\simeq\left(\frac{2^{10}e^6}{sx^2a^2}\right)\left[\frac{1-\sin^2\theta\cos^2\phi_+}{\sin^4\theta}\right]~~~.
\end{align}
The differential cross section is given by
\begin{align}
d\sigma=\frac{1}{(2\pi)^5}\frac{1}{8s^{3/2}}\left|{\cal M}\right|^2 \left|{\bf p}_+^*\right|\left|{\bf k}\right| dm_{+-} d\Omega_+^*d\Omega_k
\end{align}
where $({\bf p}_+^*,\Omega_+^*)$ is the momentum of the final positron in the final electron-positron rest frame, $({\bf k}, \Omega_k)$ is the momentum of the photon in the center-of-mass frame, and $m_{+-}^2=s^\prime=s-2\sqrt{s}k_0$.  To lowest order in $x$, the center-of-mass frame and final electron-positron frame do not differ so we will carry out the phase space integral in the center-of-mass (i.e. we expand the Jacobian to zeroth order in $x$ which is simply unity).  Then, to lowest order in $x$ and $a$, the differential cross section is
\begin{align}
d\sigma&=\frac{1}{(2\pi)^5}\frac{1}{8s^{3/2}}\left(\frac{2^{10}e^6}{sx^2a^2}\right)\left[\frac{1-\sin^2\theta\cos^2\phi_+}{\sin^4\theta}\right] \left(\frac{\sqrt{s}}{2}\right)\left(\frac{\sqrt{s}x}{2}\right)\left(\frac{\sqrt{s}}{2}\frac{dx}{\sqrt{1-x}}\right) d\Omega_+d\Omega_k
\\
&=\frac{1}{(2\pi)^5}\left(\frac{2^{4}e^6}{s a^2}\right)\left[\frac{1-\sin^2\theta\cos^2\phi_+}{\sin^4\theta}\right] \frac{dx}{x\sqrt{1-x}} d\Omega_+d\Omega_k~~~.
\end{align}
Now we write $d\Omega_+=d\cos\theta_+ d\phi_+=(2/s)dt d\phi_+$.  We also express $a^2$ in terms of $t$.  Integrating over azimuthal angles we find
\begin{align}
\frac{d\sigma}{d\cos\theta}&=-\frac{1}{(2\pi)^3}\left(\frac{2^{3}e^6}{s}\right)\left[\frac{1-\frac{1}{2}\sin^2\theta}{\sin^4\theta}\right] \frac{dx}{x\sqrt{1-x}}\frac{dt}{t}
\\
&=-\left(\frac{32\alpha^3}{s}\right)\left[\frac{1+\cos^2\theta}{(1-\cos^2\theta)^2}\right] \frac{dx}{x\sqrt{1-x}} \frac{dt}{t}~~~.
\end{align}
This will diverge if we use Eq.\ (\ref{eq:t}) to find the upper limit on the integral over $t$ by setting $\cos\theta_+=1$, i.e. $t_{\rm max}=0$.  To estimate the limits on $t$ we expand to the next nonvanishing order in $x$ and include the electron mass.  We obtain
\begin{align}
t\simeq-2p_+^0 q_+^0 \left(1-\cos\theta_+\right)-\frac{m^2 x^2}{4}\left(1+\hat{n}\cdot\hat{n}_+\right)^2\cos\theta_+~~~.
\end{align}
Setting $\left(\cos\theta_+\right)_{\rm min}=c$ and $\left(\cos\theta_+\right)_{\rm max}=1$ we find
\begin{align}
t_{\rm min}\simeq-\left(\frac{s}{2}\right) \left(1-c\right)~~~,~~~t_{\rm max}\simeq-\frac{m^2 x^2}{4}\left(1+\cos\theta\right)^2~~~.
\end{align}
Integrating over $t$ now gives us
\begin{align}
\frac{d\sigma}{d\cos\theta}&=\left(\frac{32\alpha^3}{s}\right)\left[\frac{1+\cos^2\theta}{(1-\cos^2\theta)^2}\right] \log\left[\frac{2s\left(1-c\right)}{m^2 x^2 \left(1+\cos^2\theta\right)}\right]\frac{dx}{x\sqrt{1-x}}~~~.
\end{align}
We integrate over photon energies now and find
\begin{align}
\frac{d\sigma}{d\cos\theta}&\simeq\left(\frac{32\alpha^3}{s}\right)\left[\frac{1+\cos^2\theta}{(1-\cos^2\theta)^2}\right] \log\left(\frac{k_{\rm max}}{k_{\rm min}}\right)\log\left[\frac{s^2\left(1-c\right)}{2 m^2 k_{\rm min}k_{\rm max} \left(1+\cos^2\theta\right)}\right]~~~.
\label{eq:diffxsection}
\end{align}


\end{document}